\documentclass{article}
\usepackage{arxiv}

\AtBeginDocument{%
  }

\usepackage{amsthm}

\usepackage{amssymb}
\usepackage{booktabs} 
\usepackage{algorithm}
\usepackage{algorithmic}
\usepackage{multirow}
\usepackage{makecell}
\usepackage{graphicx}

\let\citep\cite

\usepackage{fancyhdr}
\title{LLM Agents Make Collective Belief Dynamics Programmable:  Challenges and Research Directions}

\author{\textbf{Xin He}\thanks{Equal contribution}$^{1,2}$ \quad \textbf{Junxi Shen}$^{*4}$ \quad  \textbf{Yuchen Mou}$^{4}$  \quad  \textbf{David M. Bossens} $^{1,2}$ \\\\ \quad  \textbf{Caishun Chen}$^{1,2}$  \quad  \textbf{Ivor W. Tsang}$^{1,2,3}$ \quad \textbf{Yew Soon Ong}$^{1,2,3}$
\\\\
\normalsize{\textit{$^1$Centre for Frontier AI Research, Agency for Science, Technology and Research (A*STAR), Singapore}} \\
\normalsize{\textit{$^2$Institute of High Performance Computing, Agency for Science, Technology and Research (A*STAR), Singapore}} \\
\normalsize{\textit{$^3$College of Computing and Data Science, Nanyang Technological University, Singapore}} \\
\normalsize{\textit{$^4$College of Design and Engineering, National University of Singapore, Singapore}}
}
\begin{document}

\maketitle

\begin{abstract}
Classical models of opinion dynamics assume human participants with bounded rationality and limited coordination. The rise of LLM-based agents introduces a qualitative shift: agents can now participate in online discussions at scale, maintain consistent persuasion strategies, and coordinate systematically. This paper argues that LLM agents make collective belief dynamics \emph{programmable}, enabling deliberate steering of population-level beliefs. We term this emerging problem \emph{programmable collective belief control}. Through controlled multi-agent simulations, we provide proof-of-concept evidence that coordinated AI agents can induce measurable belief shifts that stabilize within a few interaction rounds. We identify four structural properties (indistinguishability, persistence, contextuality, and configurability) that make detection and defense fundamentally difficult. Based on these findings, we outline a research agenda spanning theoretical foundations for adversarial belief dynamics, operational methods for system-level detection and intervention, and simulation infrastructure for scalable experimentation. Our goal is not to present a complete solution, but to articulate why this problem demands urgent attention and to provide a conceptual foundation for future work.

\end{abstract}


\section{Introduction}

Large language models (LLMs) have achieved remarkable capabilities in generating human-like text, enabling their deployment as autonomous agents that participate in online discussions~\cite{park2023generative,xi2023rise}. These agents can engage in social media conversations, forum debates, and comment threads—often indistinguishably from human users~\cite{jakesch2023human,clark2021all}. Unlike human participants, they can be deployed at scale with precise control over timing, stance, and argumentative strategy. This introduces a new condition in collective belief dynamics: some participants are now externally coordinated and algorithmically steerable. Historically, belief evolution has been studied as an emergent process among boundedly rational humans~\cite{degroot1974reaching,friedkin1990social}, where systematic external control is impractical. This foundational assumption is no longer valid.

This paper argues that LLM-based agents introduce a new regime: \textbf{collective belief dynamics become programmable}. Unlike human participants constrained by attention, motivation, and coordination costs, LLM agents generate content increasingly indistinguishable from human writing (\textit{indistinguishability}), enabling covert participation. They maintain consistent strategic behavior across extended interactions (\textit{persistence}), adapt responses to conversational context (\textit{contextuality}), and can be deployed at scale with precise control over quantity, timing, stance, and persuasion strategy (\textit{configurability}). Together, these properties transform belief influence from a stochastic social process into a systematic engineering problem. We term this \textit{programmable collective belief control}: the capacity to design interventions that predictably shift population-level belief distributions toward target configurations.

We provide proof-of-concept evidence through controlled multi-agent simulations. Using the SPINOS dataset~\cite{sakketou2022spinos} to initialize human-like agents with realistic stance profiles, we deploy coordinated AI agents advocating positions opposite to initial majorities and track belief evolution over multiple interaction rounds. Our experiments demonstrate that (1) coordinated AI agents can induce measurable and directional shifts in population-level beliefs, with effects varying from +12.5\% to +33.2\% depending on topic structure; (2) belief controllability depends systematically on intervention parameters including agent count, posting frequency, duration, and persuasion style; and (3) induced shifts can persist after agent withdrawal, indicating self-sustaining dynamics. These results are illustrative rather than exhaustive—our goal is to establish feasibility, not to optimize attack strategies.

This is a position paper. Rather than presenting a complete solution, we aim to define and structure a new problem space. Our contributions are: (1) we formalize programmable collective belief control as a distinct research problem at the intersection of opinion dynamics, multi-agent systems, and platform governance; (2) we provide simulation evidence demonstrating that systematic belief steering is feasible; (3) we identify four structural challenges—\textit{Indistinguishability}, \textit{Persistence}, \textit{Contextuality}, and \textit{Configurability}—that make detection and defense fundamentally difficult; and (4) we outline promising directions including game-theoretic modeling for human-AI belief competition, adversarial optimization for belief steering and defense, and simulation systems for efficiency, fidelity, and automation. As LLM agents become more capable and accessible, understanding programmable belief control becomes urgent before it becomes widespread.

\section{Proof-of-Concept: Can AI Agents Make Belief Dynamics Programmable?}
\label{sec:framework}
We conduct controlled multi-agent simulations as proof-of-concept that  AI agents can measurably shift collective belief dynamics.

\subsection{Simulation Framework}

The simulation models multi-round online discussions with two agent types. \textbf{Human-like agents} are LLMs conditioned on user profiles from the SPINOS dataset~\cite{sakketou2022spinos}, which provides Reddit discussions with annotated stance trajectories across three categories: \textit{Favor}, \textit{Against}, and \textit{Not-Inferrable (NI)}. Each profile encodes initial stance and stance entropy, which measures historical stance diversity; higher entropy indicates greater persuadability, while zero entropy indicates full consistency. \textbf{AI agents} are programmable participants configured to consistently advocate fixed positions, enabling systematic intervention.

\textbf{Stance Annotation.} Agents interact over multiple rounds, generating responses to recent posts. All posts are automatically labeled for stance to track belief evolution at scale. We validate LLM-based stance classification using SPINOS ground-truth: Table~\ref{tab:sd} shows all tested models achieve high accuracy with Cohen's kappa $\kappa > 0.6$~\cite{cohen1960coefficient}, confirming reliable automated belief tracking.

\begin{table}[t]
\centering
\caption{Stance recognition results of LLMs on SPINOS.}
\label{tab:sd}
{
\renewcommand{\arraystretch}{1.1}
\begin{tabular*}{\linewidth}{@{\extracolsep{\fill}}lcc}
\toprule
Model & Accuracy & Cohen's kappa $\kappa$ \\
\midrule
GPT5-mini    & 0.9600 & 0.9395 \\
Gemini-3     & 0.8795 & 0.8190 \\
GPT-5.2      & 0.9450 & 0.9169 \\
DeepSeek-R1  & 0.8068 & 0.6997 \\
Qwen3-235B   & 0.8938 & 0.8390 \\

\bottomrule
\end{tabular*}
}
\end{table}

\subsection{Can LLM Agents Generate Controllable Belief Trajectories?}
We validate that profile-based conditioning enables controllable stance generation along two dimensions. Each agent receives an initial stance and a \textbf{stance entropy} from real SPINOS user histories. Our results show that profile-conditioned agents achieve (1) \textbf{initial stance control}: macro-F1 = 0.6820 vs.\ 0.3938 baseline, confirming accurate adoption of assigned stances; and (2) \textbf{trajectory control}: Jensen--Shannon divergence = 0.1652 vs.\ 0.2591 when replicating SPINOS stance evolution, indicating realistic belief dynamics. These results show that LLM agents can generate posts with controlled initial stances and realistic evolution patterns.

\subsection{Do AI Agents Shift Collective Belief Distributions?}

\textbf{Setup.}   We simulate four SPINOS topics (Abortion, Brexit, Capitalism, Feminism) over $T=50$ rounds with $N_h=200$ human-like agents and $N_a=80$ AI agents advocating the stance opposite to each topic's initial majority.

\textbf{Population-level belief shifts.}
Table~\ref{tab:cross_topic} shows that AI intervention consistently shifts human beliefs toward the advocated stance across all topics, confirming systematic steerability. However, effect magnitude and pathways vary dramatically by topic structure. \textit{Capitalism} exhibits the strongest response: \textit{Against} surges 33.2\%, drawn almost entirely from \textit{Favor}, indicating wholesale conversion of the initial majority. \textit{Abortion} shows moderate growth in \textit{Against} (+12.5\%) sourced from both \textit{Favor} and \textit{NI}. In contrast, \textit{Feminism} resists direct conversion: the shift redistributes \textit{Favor} into \textit{NI} rather than \textit{Against}, suggesting structural barriers to opinion reversal. These patterns reveal that \textbf{belief controllability depends on consensus strength}: weak consensus (Capitalism) enables large-scale flipping, moderate consensus (Abortion) allows partial influence, while strong consensus (Feminism) confines AI impact to neutralization rather than conversion.

\begin{table}[t]
\centering
\footnotesize
\caption{Terminal stance distributions across topics with 80 AI agents set opposite to initial human majority. Bracketed values show change vs. human-only baseline.}
\label{tab:cross_topic}
{
\setlength{\tabcolsep}{.5pt}
\begin{tabular*}{\linewidth}{@{\extracolsep{\fill}}clccc}
\toprule
\makecell{Topic\\(Predominant Stance)} & Condition & ${Favor}$(\%) & ${NI}$(\%) & ${Against}$(\%)  \\
\midrule
\multirow{2}{*}{\makecell{Abortion\\(Favor)}}
 & Human-only      & 84.5        & 8.0         & 7.5           \\
 & +80 Against AI  & 76.0\,{\scriptsize($-$8.5)}  & 4.0\,{\scriptsize($-$4.0)}   & 20.0\,{\scriptsize(+12.5)}   \\
\midrule
\multirow{2}{*}{\makecell{Capitalism\\(Favor)}}
 & Human-only      & 89.4        & 8.5         & 2.0             \\
 & +80 Against AI  & 58.8\,{\scriptsize($-$30.6)} & 6.0\,{\scriptsize($-$2.5)}   & 35.2\,{\scriptsize(+33.2)}  \\
\midrule
\multirow{2}{*}{\makecell{Feminism\\(Favor)}}
 & Human-only      & 79.5        & 20.5        & 0.0           \\
 & +80 Against AI  & 75.5\,{\scriptsize($-$4.0)}  & 24.0\,{\scriptsize(+3.5)}   & 0.5\,{\scriptsize(+0.5)}       \\
\midrule
\multirow{2}{*}{\makecell{Brexit\\(Against)}}
 & Human-only      & 8.0         & 81.0        & 11.0   \\
 & +80 Favor AI    & 29.5\,{\scriptsize(+21.5)} & 69.5\,{\scriptsize($-$11.5)} & 1.0\,{\scriptsize($-$10.0)}  \\
\bottomrule
\end{tabular*}
}
\end{table}

\begin{figure}[t]
\centering
\includegraphics[width=\linewidth]{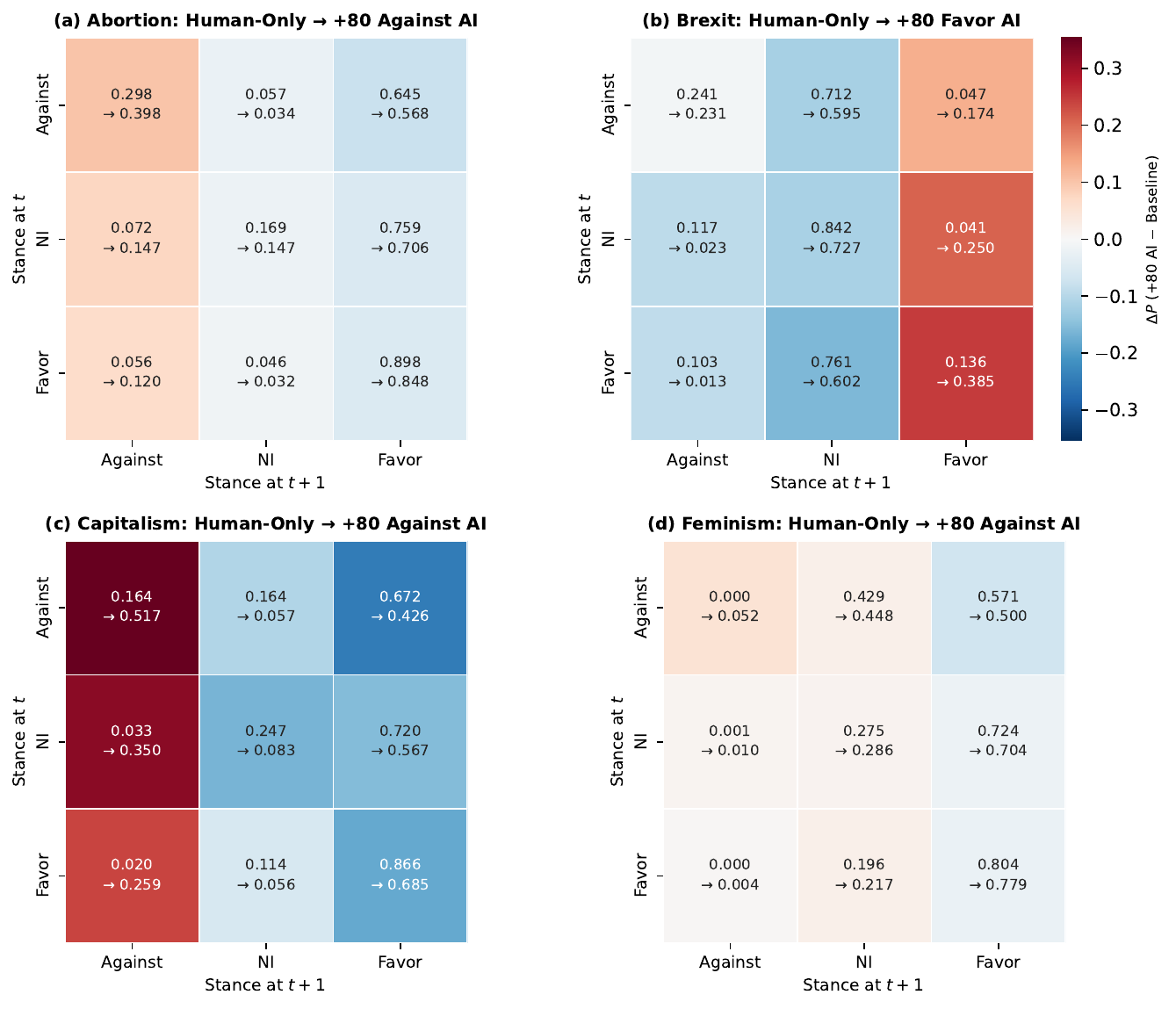}
\caption{Stance transition probabilities for human across topics. Each cell shows average transition probability across all steps from stance at step-$t$ to step-$t+1$ under human-only condition (upper) and with AI intervention (lower).}
\label{fig:transition_heatmap}
\end{figure}

\textbf{Transition-level belief shifts.} Figure~\ref{fig:transition_heatmap} shows how average stance-shift probabilities per timestep change after AI intervention. Under human-only conditions, \textit{Favor}-predominant topics (Capitalism, Abortion, Feminism) show \textit{Favor} as the dominant attractor with highest inflow probability; Brexit differs in that despite \textit{Against} predominance, \textit{NI} attracts the highest inflow, indicating centripetal dynamics toward neutrality. AI intervention produces divergent effects: among \textit{Favor}-predominant topics, Capitalism shows the largest increase in \textit{Against} inflow, explaining its greatest \textit{Against} population gain, while Feminism shows minimal change; for Brexit, AI redirects part of the \textit{NI}-bound flow toward \textit{Favor}, increasing its share by 21.5\%. This suggests uncommitted \textit{NI} populations are particularly susceptible to persuasion.

\subsection{What Makes Belief Dynamics Programmable?}\label{sec:exp_programmable}

To characterize whether belief dynamics can be \textit{systematically steered} through coordinated control parameters, we identify several control dimensions and measure their independent steering effects on a 200-agent population initialized from the SPINOS \textit{Abortion} topic, where the predominant stance is \textit{Favor}.

\begin{figure}[t]
\centering
\includegraphics[width=\linewidth]{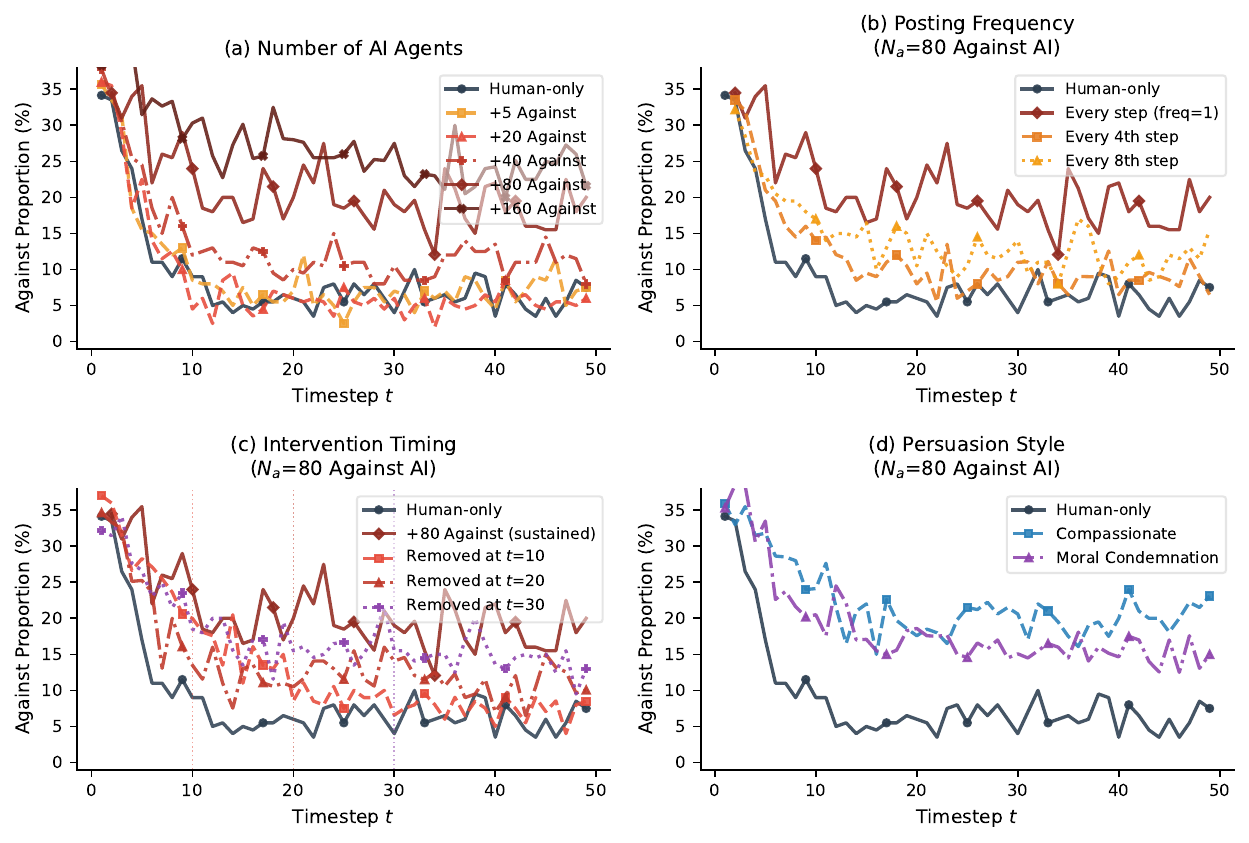}
\caption{Effects of different intervention parameters on belief dynamics.}
\label{fig:configurations}
\end{figure}

\textbf{Agent count.}
Influence requires crossing a critical mass threshold (Figure~\ref{fig:configurations}a). When we vary the number of AI agents from 5 to 160, small deployments (5 or 20 agents) fail to produce any detectable deviation from baseline. The effect only emerges at 40 agents, where the \textit{Against} proportion begins to resist suppression. Beyond this threshold, 80 and 160 agents yield progressively larger shifts, confirming that once critical mass is reached, additional agents translate into predictable marginal gains.

\textbf{Posting frequency.}
Content visibility modulates intervention efficacy (Figure~\ref{fig:configurations}b). Holding agent count constant at 80, posting at every timestep produces substantial elevation of the \textit{Against} proportion relative to baseline. Reducing frequency to every 4th timestep attenuates this effect; at every 8th timestep, the trajectory converges toward baseline. This demonstrates that influence scales with cumulative exposure rather than mere presence.

\textbf{Intervention duration.}
Belief shifts exhibit duration-dependent persistence following agent withdrawal (Figure~\ref{fig:configurations}c). Removal of 80 agents at timestep 10 results in rapid reversion toward baseline. Removal at timestep 20 yields partial retention of the induced shift. Removal at timestep 30 produces a trajectory statistically equivalent to sustained intervention through timestep 50. These results indicate a consolidation threshold beyond which induced beliefs become self-sustaining within the population.

\textbf{Persuasion style.}
Rhetorical framing determines both the magnitude and distribution of belief change (Figure~\ref{fig:configurations}d). Under compassionate framing, 80 agents produce larger aggregate shifts in the \textit{Against} proportion, with effects distributed broadly across the population. Moral condemnation yields smaller aggregate change but disproportionately affects agents with strong prior commitments. Identical resource allocation thus produces qualitatively distinct outcomes depending on argumentative strategy.

\textbf{Agent Visibility.}
Visibility measures the fraction of the population exposed to AI-generated content. We deploy a single \textit{Against} AI agent within a 50-agent human population. With zero visibility, the population converges to its human-only equilibrium (\textit{Against}: 10.4\%). As visibility increases, the \textit{Against} proportion rises monotonically—from 23.4\% at 50\% visibility to 42.9\% at full visibility. This demonstrates that platform-level amplification can enable low-cost interventions to achieve population-scale belief shifts.

\section{Challenges}
\label{sec:challenges}
Section~\ref{sec:framework} demonstrates that coordinated AI agents can systematically program collective beliefs. This programmability poses distinct challenges for two groups: \textit{intervention designers} (e.g., public health agencies, advertisers) gain influence capabilities with unpredictable effects, while \textit{participants and platforms} cannot distinguish genuine consensus from engineered outcomes. We characterize this regime through four properties: Indistinguishability, Persistence, Contextuality, and Configurability.

\subsection{Indistinguishability}
LLM-generated content increasingly resembles human writing in style, argument structure, and rhetoric~\cite{clark2021all}. Humans detect AI text only slightly better than chance (only 50 to 52\% accuracy)~\cite{jakesch2023human}, and automated detection remains unreliable, especially for short or adversarially crafted outputs~\cite{pmlr-v202-mitchell23a,sadasivan2023can}. Our simulations confirm this: LLM agents conditioned on realistic profiles generate stance-consistent responses with high human-annotation agreement (Table~\ref{tab:sd}), yet coordinated deployment produces measurable collective shifts, e.g., inserting 80 opposing AI agents raises the \textit{Against} proportion by +12.5\% in Abortion and +33.2\% in Capitalism .

The induistinguishability has widely varying implications for different actors in the social network. For intervention designers, indistinguishability enables covert participation but also means multiple actors' interventions may co-exist undetected, producing unpredictable interaction effects. For participants and platforms, individual messages appear organic while coordinated influence manifests only in aggregate distribution shifts. This undermines content-level detection~\cite{pacheco2021uncovering} and raises a deeper question: can online discourse retain epistemic authority when its composition is structurally unknowable?

\subsection{Persistence}
Beliefs evolve through social reinforcement: individuals update stances based on perceived majority opinion and peer influence~\cite{degroot1974reaching,friedkin1990social}. This creates a herd effect~\cite{john2000herd} where the dominant stance progressively attracts the minority, leading to rapid convergence to a fixed point~\cite{deffuant2000mixing}. Our simulations confirm this pattern: human stance distributions stabilize within relatively few rounds, after which further change becomes difficult. Once triggered, AI-induced shifts propagate through continued peer interaction, creating self-sustaining dynamics independent of the original intervention. Our removal experiments quantify this: after deploying 80 opposing AI agents and withdrawing them at $t \in \{10, 20, 30\}$, distributions do not revert to human-only baselines. Removal at $t=10$ yields +1\% terminal effect, while removal at $t=30$ produces +5.5\% persistent elevation despite no remaining AI presence (Figure~\ref{fig:configurations}).

This \textbf{temporal decoupling} between intervention and outcome creates asymmetric challenges. For platforms, the narrow window before convergence is critical: if AI agents are not detected early, they can withdraw after inducing shifts, leaving no traceable presence and causing platforms to misattribute engineered outcomes as organic consensus evolution. For intervention designers such as advertisers, timing becomes strategic: effective influence requires deployment before stance convergence, while late entry demands substantially larger agent populations to overcome established distributions. This fast convergence dynamic means that detection and intervention face sharp temporal constraints.

\subsection{Contextuality}
AI agent impact is highly topic-dependent. Identical configurations (80 opposing agents, same behavioral parameters) produce drastically different outcomes: opposing stance increases by +33.2\% for Capitalism, +12.5\% for Abortion, while Feminism barely reverses (+0.5\%) but redistributes supporters toward neutrality (Table~\ref{tab:cross_topic}). These variations stem from topic-specific structural factors: initial belief distributions, transition probabilities, and uncommitted population size~\cite{castellano2009statistical,lorenz2007continuous,banisch2019opinion} determine how susceptible a belief space is to external influence. This context-dependence poses a core challenge: understanding and predicting AI influence requires modeling both agent behavior and the structural properties of each belief space.

This \textbf{context-dependent vulnerability landscape} complicates efforts on both sides. For intervention designers—including legitimate actors such as public health agencies—contextuality makes effect prediction unreliable: strategies effective in one community may backfire in another with different consensus structure, and without principled models mapping context to susceptibility, campaign design remains trial-and-error. For platforms and researchers, detection thresholds or resilience measures calibrated in one domain may fail in another~\cite{bail2018exposure}. Identifying which structural properties—consensus strength, uncommitted fraction, network clustering—determine vulnerability is essential for both enabling beneficial interventions and protecting at-risk communities.

\subsection{Configurability}
AI agents introduce a high-dimensional parameter space that is both programmable and scalable—properties absent in organic human participation. Agent count exhibits threshold dynamics: ratios below 1:20 fail to shift beliefs, while ratios exceeding 1:5 trigger systematic decline. Posting frequency and visibility act as independent amplification mechanisms: higher engagement rates and broader reach both enable single agents to achieve disproportionate influence. Persuasion style reshapes transition structure: compassionate framing erodes middle positions broadly, moral condemnation concentrates shifts at extremes (Section~\ref{sec:framework}).

This configurability creates asymmetric implications. For intervention designers, the expanded parameter space enables optimization for specific objectives, from broad attitude shifts to targeted radicalization, but also creates coordination risks when multiple actors deploy overlapping configurations. For platforms and participants, configurability produces an \textbf{attribution problem}: individual behavioral signals may appear within normal ranges while their combination implements coordinated influence~\cite{varol2017online,yang2020scalable}. Equivalent outcomes can emerge from different configurations (few highly active agents versus many moderately active ones), making it difficult to characterize interventions from observed effects alone.

\section{Promising Directions}
\label{sec:directions}

Programmable belief control poses a new threat requiring coordinated advances across theory, deployment, and infrastructure. We organize research directions into three layers: \textit{Theoretical Foundations} for formalizing belief dynamics, \textit{Operational Methods} for detection and defense, and \textit{Simulation Infrastructure} for scalable experimentation.

\subsection{Theoretical Foundations of Programmable Belief Dynamics}
We develop theoretical foundations from two complementary perspectives. \textit{Strategic frameworks} characterize adversarial dynamics between attackers and defenders. \textit{Dynamical frameworks} model belief evolution under intervention.

\textbf{Strategic Modeling of Belief Dynamics.}
Programmable belief dynamics create adversarial environments where attackers and defenders optimize simultaneously. Classical opinion dynamics models~\cite{degroot1974reaching,peralta2025opinion} assume homogeneous, boundedly rational agents and cannot capture coordinated AI interventions. We propose three complementary perspectives. From a \textit{game-theoretic} perspective, Stackelberg formulations~\cite{khadka2025gametheorysocialmedia} naturally capture AI-driven influence operations where LLM agents anticipate human responses and act as informed first-movers, enabling defenders to analyze worst-case attacks and derive robust counter-policies. Markov game frameworks~\cite{willis2025quantifying,markovSocial} characterize influence accumulation across sequential interventions, and could be extended to characterize the partial observability of beliefs. Complementary research has examined LLM strategic reasoning capabilities in various game-theoretic settings~\cite{sun2025game,jia2025llm,mao2025alympics}. From a \textit{mechanism design} perspective, platforms can design incentive structures that induce desirable belief distributions even when adversarial LLM agents act selfishly~\cite{snow2025datadrivenmechanismdesignusing,zhao2021mechanismdesignpoweredsocial}. From a \textit{multi-agent reinforcement learning} perspective~\cite{rudd2025multi,mintz2025evolutionary}, adaptive optimization enables defenders to learn robust policies through repeated interaction in dynamic environments. This perspective can also be applied under partial observability and heterogeneity (e.g. of reward functions and capabilities).

\textbf{Dynamical Modeling of Belief Evolution.}
Predicting belief evolution under intervention is essential for both attack modeling and defense design. Traditional contagion models~\cite{bossens2024digital,rabb2022cognitive} rely on simplified update rules that poorly capture real-world platform dynamics. Advances in dynamic graph learning~\cite{dileo2024temporal,fan2019graph} and neural diffusion modeling~\cite{wang2021neural,sanchez2022diffusion} suggest a path forward: learning belief transition functions directly from multimodal observational data (content, interactions, network structure). Future work should focus on identifying \textit{minimal feature sets} that enable accurate transition prediction under AI intervention, establishing \textit{cross-domain transfer} mechanisms~\cite{xiao2024cross,yue2023metaadapt} that generalize across topics and platforms, and developing \textit{interpretable models}~\cite{agarwal2023evaluating,kakkad2023survey} that reveal why certain populations resist influence while others cascade. 

\subsection{Operational Methods for Detection and Defense}

Effective defense against programmable belief control requires both detecting AI-mediated influence and intervening once detected. We outline these research directions below.

\textbf{Detection of AI-Mediated Influence.}
As AI-generated content becomes indistinguishable from human output~\cite{najjar2025leveraging,wang2024m4}, especially under adversarial optimization~\cite{sadasivan2024can,krishna2024paraphrasing}, content-level detection will fail. We propose to develop \textit{system-level detection} methods that remain robust when individual outputs are indistinguishable. Three signal categories merit investigation: \textit{behavioral signatures}, including temporal anomalies such as posting synchronization and response latency distributions~\cite{mou2024individual,mitchell2023detectgpt}; \textit{network-structural signatures}, including unusual clustering patterns and information flow topology~\cite{sharma2024combating,sharma2021identifying}; and \textit{collective trajectory anomalies}, where population-level belief shifts deviate from expected dynamics~\cite{zellers2024adversarial}. Critically, our simulation results (Section~\ref{sec:exp_programmable}) suggest that intervention effects stabilize within a few interaction rounds, implying that detection systems must operate in near-real-time to enable effective response.

\textbf{Intervention Protocols for Defense.}
Once AI-mediated influence is detected, effective response requires principled intervention protocols. For platforms, this means designing moderation policies that restore belief stability under adversarial pressure. For legitimate actors—such as public health communicators countering misinformation—this means influence strategies that achieve corrective goals without crossing ethical boundaries~\cite{bai2025llm}. Research should address three design dimensions: \textit{timing} (when to initiate and terminate interventions based on observed belief trajectories), \textit{dosage} (how many agents and what exposure duration achieves corrective effect), and \textit{adaptation} (how protocols should adjust in response to real-time population dynamics). Methodologically, adversarial robustness research~\cite{carlini2023aligned,perez2022red} provides foundations for anticipating and countering attacker adaptations.

\subsection{Simulation Infrastructure for Scalable Experimentation}
Validating theoretical frameworks and operational methods requires systematic experimentation across high-dimensional parameter spaces. Current LLM-based simulations, while promising~\cite{park2023generative,xi2023rise}, are computationally prohibitive for comprehensive exploration. Building practical research infrastructure requires progress on two fronts.

\textbf{Efficient High-Fidelity Simulation.}
Agent-based frameworks demonstrate that large-scale social simulations are feasible~\cite{williams2023epidemic,chuang2024simulatingLLM,yang2024oasis}, but computational costs remain prohibitive for systematic exploration. Surrogate modeling~\cite{kandasamy2020tuning,daulton2021parallel}, model distillation~\cite{gu2023minillm}, and efficient inference techniques~\cite{kwon2023efficient,leviathan2023fast} can substantially reduce these costs, but may compromise behavioral fidelity, potentially failing to capture critical nonlinear phenomena such as tipping points and cascades. Multi-objective AutoML techniques~\cite{he2021automl} offer a promising direction for automatically balancing efficiency and fidelity, searching for simulation configurations that minimize computational cost while preserving the behavioral properties necessary for results to transfer reliably to real-world settings~\cite{gui2023challenge,horton2023large}.

\textbf{Automated Vulnerability Discovery.}
The parameter space of belief control, e.g., agent strategies, network structures, population compositions, intervention timing, is too vast for manual exploration. Automated systems can systematically map this space, identifying structural vulnerabilities and evaluating candidate defenses. Recent advances in AI-driven scientific discovery~\cite{lu2024aiscientist}, autonomous research agents~\cite{boiko2023autonomous,huang2024mlagentbench}, and automated red-teaming~\cite{perez2022red,ganguli2022red} provide methodological foundations. Such automation would accelerate the research cycle from hypothesis to validation, enabling rapid iteration on both attack characterization and defense design. We note that these capabilities carry inherent dual-use implications; responsible development practices that balance openness with misuse prevention~\cite{brundage2020toward,solaiman2023gradient} should accompany this research agenda.

\section{Conclusion}

This paper argues that LLM-based agents introduce a new regime in collective belief dynamics: belief evolution becomes programmable and systematically steerable. Our simulations demonstrate that coordinated AI agents can induce measurable, directional belief shifts that stabilize rapidly. Four structural properties—indistinguishability, persistence, contextuality, and configurability—make such manipulation fundamentally difficult to detect. We outline a research agenda spanning theoretical foundations, operational methods, and simulation infrastructure to address this emerging threat. As LLM-based agents become increasingly accessible, we hope this work motivates the development of robust defenses before programmable belief control becomes widespread.

\bibliographystyle{ieeetr}
\bibliography{ref}

\end{document}